\input harvmac.tex
\def\scrn{{\cal{N}}}
\def\tilde{\widetilde}
%%%%%%%%%%%%%%%%%%%%%%%%%%%%%%%%%%%%%%%%%%%%%%%%%%%
\Title{\vbox{\baselineskip12pt\hbox{hep-th/9707158}
\hbox{IAS-97-84, RU-97-62}}}
{\vbox{\centerline{The Coulomb Branch of (4,4) Supersymmetric}
\centerline{}
\centerline{Field Theories in Two Dimensions}}}

\centerline{Duiliu-Emanuel Diaconescu}
\smallskip
\smallskip
\centerline{Department of Physics and Astronomy}
\centerline{Rutgers University }
\centerline{Piscataway, NJ 08855-0849, USA}
\smallskip
\smallskip
\smallskip
\centerline{and}
\smallskip
\centerline{Nathan Seiberg}
\smallskip
\smallskip
\centerline{Institute for Advanced Study}
\centerline{Princeton, NJ 08540, USA}
\bigskip
\bigskip
\noindent
We study (4,4) supersymmetric field theories in two dimensions with
a one dimensional Coulomb branch.  These theories have
applications in string theory.  Our analysis explains the known
relation between $A-D-E$ groups and modular invariants of affine
$SU(2)$. 

%\draftmode

\Date{July 1997}

\newsec{Introduction}

Supersymmetric theories in various dimensions can be analyzed exactly.
This has led to new insights in field theory and string theory.  In
particular, some theories with eight supercharges can be studied as
toroidal compactifications of the minimal supersymmetric theories in
six dimensions.  This was done explicitly for compactifications to 5,
4 and 3 dimensions (see
\ref\gms{O.J. Ganor, D.R. Morrison, N. Seiberg, Nucl.Phys. {\bf B487}
(1997) 93, hep-th/9610251.}
and references therein).  The purpose of this paper is to extend this
discussion to compactification to 2 dimensions.

These field theories occur in string theory in different setups.
First, they occur in the matrix model of M theory
\ref\bfss{T. Banks, W. Fischler, S. H. Shenker, L. Susskind,
Phys.Rev. {\bf D55} (1997) 5112, hep-th/9610043.}.
Second, they occur in compactifications on singular manifolds.
Finally, they occur on various branes.  The six dimensional theories
are the low energy theories on the 5-branes of the heterotic theory.
They can be viewed as systems of 5-branes and parallel 9-branes in M
theory
\ref\eeh{M.J. Duff, R. Minasian, E. Witten, Nucl.Phys. {\bf B465}
(1996) 413,  hep-th/9601036; O.J. Ganor, A. Hanany, Nucl.Phys. {\bf
B474} (1996) 122, hep-th/9507036; N. Seiberg, E. Witten,
Nucl.Phys. {\bf B471} (1996) 121, hep-th/9603003.}
or in IIB theory
\ref\dbrane{ J. Polchinski, S. Chaudhuri, C.V. Johnson,
hep-th/9602052; J. Polchinski, hep-th/9611186.}.
Then one can think of the background
9-branes as breaking 16 of the supersymmetries and the 5-branes can be
viewed as probes
\ref\probe{M.R. Douglas, G. Moore, hep-th/9603167; M.R. Douglas,
 hep-th/9604198.}.  
Similarly, the five dimensional theories occur
on 4-branes probing a background with 8-branes
\ref\fived{N. Seiberg, Phys.Lett. {\bf B388} (1996) 753, 
hep-th/9608111.},
the four dimensional theories occur
on 3-branes probing a background with 7-branes
\ref\bds{T. Banks, M.R. Douglas, N. Seiberg, Phys.Lett. {\bf B387}
(1996) 278, hep-th/9605199.},
and the three dimensional theories occur
on 2-branes probing a background with 6-branes
\ref\IR{N. Seiberg, Phys.Lett. {\bf B384} (1996) 81, hep-th/9606017.}.
Here we extend it to a 1-brane probing a background with 5-branes.
Some aspects of this system have been discussed before
\ref\chs{C.G. Callan, J.A. Harvey, A. Strominger, Nucl.Phys. {\bf
B359} (1991) 611; C.G. Callan, J.A. Harvey, A. Strominger,
hep-th/9112030; S.-J. Rey, In The  Proc. of the Tuscaloosa Workshop
1989, 291; Phys. Rev. {\bf D43} (1991) 526; S.-J. Rey, In DPF Conf.
1991, 876.}
Finally, the system of 0-branes probing a
background of 4-branes was studied in
\ref\dkps{M.R. Douglas, D. Kabat, P. Pouliot, S. Shenker,
Nucl.Phys. {\bf B485} (1997) 85, hep-th/9608024.}.

We limit ourselves to theories with a one dimensional
Coulomb branch.  These theories are classified by an $A-D-E$ type
classification.  The $A_{N_f-1}$ theories correspond to $U(1)$ gauge
theories with $N_f$ electrons (their global symmetry includes an
$SU(N_f)$ factor, hence the term $A_{N_f-1}$ theories).  They
correspond to a single $d$ dimensional probe in the presence of $N_f$
$d+4$ dimensional background branes.  The $D_{N_f}$ theories
correspond to $SU(2)$ gauge theories with $N_f$ quarks (their global
symmetry includes an $SO(2N_f)$ factor, hence the term $D_{N_f}$
theories).  They correspond to a single $d$ dimensional probe in the
presence of $N_f$ $d+4$ dimensional background branes and a $d+4$
dimensional orientifold.  The other theories, including the various
$E$ theories correspond to background branes at strong coupling.

There is a crucial difference between the higher dimensional field
theories and the two or one dimensional cases.  Here, the notion of
moduli space of vacua is ill defined.  One should integrate over all
the vacua.  However, one intuitively expects, in the spirit of
the Born-Oppenheimer approximation, a notion of a low energy theory
obtained by integrating out the high energy modes, which looks like a
non-linear sigma model on a target space.  We will refer to this
target space as the moduli space of vacua.

As in the higher dimensional analogs, these systems have a Coulomb
branch of vacua corresponding to motion of the probe away from the
background branes.  The Higgs branch corresponds to the absorption of
the probe in the background branes as an instanton
\ref\dougin{M.R. Douglas, hep-th/9512077.}.
The theory on this branch was used recently in
\nref\aharo{O. Aharony, M. Berkooz, S. Kachru, N. Seiberg,
E. Silverstein, hep-th/9707079.}%
\nref\wittenhiggs{E. Witten, hep-th/9707093.}%
\refs{\aharo,\wittenhiggs} to give a description of the (2,0) field
theory
\nref\separate{E. Witten, hep-th/9707121, Strings '95 (World
Scientific, 1996).}%
\nref\strominger{A. Strominger, Phys.Lett. {\bf 383B} (1996) 44,
hep-th/9512059.}%
\refs{\separate,\strominger} and string theory
\ref\ncst{N. Seiberg, hep-th/9705221.}.  
As in
\ref\nonren{N. Seiberg, Phys. Lett. {\bf B318} (1993) 469,
hep-ph/9309335.},
we can promote the gauge coupling to a background superfield.  In our
case the gauge coupling is in a vector superfield
\ref\aps{P. C. Argyres, M. R. Plesser, A. Shapere,
Phys. Rev. Lett. {\bf 75} (1995) 1699, hep-th/9505100.},
and therefore the Higgs branch is not corrected by quantum effects.
The Coulomb branch can be corrected. 

Here we will focus on theories with a one dimensional Coulomb branch.
Its metric turns out to be completely determined by the global
symmetries of the problem and a simple one loop computation.  This
metric leads to two separate problems.  First, in most cases the
Coulomb branch develops an infinitely long tube.  It is not completely
clear how to interpret the physics along the tube.  More importantly,
in some cases (e.g.\ in an $SU(2)$ gauge theory without quarks) the
metric has a singularity at a finite distance on the moduli space
beyond which the metric ceases to be Riemannian.  This situation is
reminiscent of similar problems in four
\nref\swi{N. Seiberg and E. Witten, Nucl.Phys. {\bf B426} (1994) 19,
 hep-th/9407087.}%
\nref\swii{N. Seiberg and E. Witten, Nucl.Phys. {\bf B431} (1994) 484,
 hep-th/9408099.}%
\refs{\swi,\swii} and three
\nref\swthree{N. Seiberg and E. Witten, In: {\it The Mathematical
Beauty of Physics, A Memorial Volume for Claude
Itzykson,} J.M. Drouffe and J.B. Zuber, Eds., World Scientific,
1997, hep-th/9607163.}%
\refs{\IR,\swthree} dimensions.  In both cases the problem is solved
by realizing that the degrees of freedom used to describe the Coulomb
branch are not appropriate and one needs to dualize them.  In terms of
the dual degrees of freedom there exists a Riemannian metric
everywhere on the moduli space.  

We find the generation of the infinitely long tube counter intuitive.
It means that the wave function of the light modes is not normalizable
near the origin (it is also not normalizable far out along the flat
direction).  This is confusing because classically this region is
compact.  Therefore, we speculate that there exists another
description of the field theory near the origin, which is compact
there in the sense that the wave function is normalizable.  Some
evidence for this speculation comes from the second problem mentioned
above which suggests that in some, yet to be found, ``dual variables''
there is no problem with a non-Riemannian metric.  In the following
sections we will consider the compactifications of these theories from
three to two dimensions and will find further evidence for this
speculation.

In the context of string theory we find these theories on D1-branes near
D5-branes and orientifolds.  Therefore, we find them in IIB
compactifications on $T^4/Z_2$.  A careful examination of the action of
the $Z_2$ shows that this theory is dual to a compactification of IIA on
K3
\ref\senkt{A. Sen, Nucl.Phys. {\bf B474} (1996) 361, hep-th/9604070;
A. Sen, hep-th/9609176; J. Harvey, talk given at the Jerusalem Winter
School on Strings and Duality, Jan. 1997.}.  
The Coulomb branch of the 1+1 dimensional theory is then mapped, after
T-duality, to the underlying K3.  This space does not have any long
tubes.  Instead, its singularities are classified by $A-D-E$.
Therefore, it is reasonable to expect that in some dual variables the
region around the origin can be replaced by an ALE space with an
$A-D-E$ singularity.  Such a proposal has already been made in
\ref\ogva{H. Ooguri, C. Vafa, Nucl.Phys. {\bf B463} (1996) 55,
hep-th/9511164; D. Kutasov, Phys. Lett {\bf B383} (1996) 48,
hep-th/9512145; H. Ooguri and C. Vafa, hep-th/9702180.}
(see however,
\ref\Anselmi{D. Anselmi, M. Bill\'o, P. Fr\`e, L. Girardello,
A. Zaffaroni, Int.J.Mod.Phys. {\bf A9} (1994) 3007, hep-th/9304135.}).
It should be stressed that the 1+1 dimensional theory on the
ALE space is not merely an orbifold theory because the $\theta$ angle
there vanishes rather than being equal to $\pi$
\ref\aspinwall{P. Aspinwall, Phys. Lett. {\bf B357} (1995) 329,
hep-th/9507012.}. 

In section 2 we present the general formalism for describing these
systems.  In section 3 we study $U(1)$ gauge theories with $N_f$
electrons and in section 4 we study $SU(2)$ gauge theories with $N_f$
quarks.  In section 5 we discuss three theories, which do not have a
simple gauge theory Lagrangian, whose global symmetries are
$E_{6,7,8}$.  In section 6 we focus on the region in the moduli space
near the origin -- the tube -- and present an explanation of the known
connection between $A-D-E$ symmetries and the classifications of
modular invariants of $\widehat {SU(2)}$
\ref\ciz{A. Cappelli, C. Itzykson, J.B. Zuber, Nucl. Phys. {\bf B280}
(1987) 445, Comm. Math. Phys. {\bf 113} (1987) 1.}.

\newsec{Preliminaries}

The $\scrn=(4,4)$ supersymmetry algebra in two dimensions is the
dimensional reduction of the six dimensional $\scrn=(1,0)$.  The
supersymmetry charges of the latter are symplectically real chiral
spinors transforming as the $\bf{4}$ of $Spin(1,5)$ and as the
doublet of the R-symmetry group $SU(2)_R$.  Dimensional reduction to
two dimensions yields eight real supercharges in
\eqn\suprep{\eqalign{
({\bf 2},{\bf 1},{\bf 2})^+\oplus ({\bf 1},{\bf 2},{\bf
2})^- \qquad {\rm of} \qquad
& Spin(4)\times SU(2)_R \times SO(1,1)\cr
&\subset Spin(1,5)\times
SU(2)_R\cr}}
(where the representations of $Spin(4)$ are labeled by their two
$SU(2)$ subgroups).
As in six dimensions, this algebra has two massless representations: a
vector multiplet and a hypermultiplet.  In two dimensions these two
representations are related by duality (more about this below).

A hypermultiplet contains four scalars transforming as $({\bf 1},{\bf
1},{\bf 2}) \oplus({\bf 1},{\bf 1},{\bf 2}) $ under $Spin(4)\times
SU(2)_R $ and four fermions.  In terms of an $\scrn=(2,2)$ subalgebra
the hypermultiplet decomposes into two chiral multiplets $\Phi$ and
$\tilde \Phi$ satisfying
\eqn\constrh{\eqalign{
&\bar D_{+}\Phi=0,\qquad \bar D_{-}\Phi=0 \cr
&\bar D_{+}\tilde \Phi=0,\qquad \bar D_{-}\tilde \Phi=0.\cr}}
The scalars in these multiplets parameterize a ``Higgs branch,'' which
is a hyper-K\"ahler manifold.  The three complex structures are in the
$\bf 3$ of $SU(2)_R$.  Along the Higgs branch there are
scalars transforming in ${\bf 3}\oplus {\bf 1}$ of $SU(2)_R$.  The
$Spin(4)$ global symmetry does not act on the Higgs branch but it acts
on the fermions.

The vector multiplet is also known as a twisted multiplet.  In terms
of an $\scrn=(2,2)$ subalgebra it decomposes into a chiral multiplet
$\Phi$ and a twisted chiral multiplet $\Lambda$ satisfying
\eqn\constr{\eqalign{
&\bar D_{+}\Phi=0,\qquad \bar D_{-}\Phi=0 \cr
&D_{+}\Lambda=0,\qquad \bar D_{-}\Lambda=0 .\cr}}
The four real scalars contained in these superfields represent the
components of the six dimensional vector field $A_{\mu}$ along the
reduced directions.  They transform as $({\bf 2},{\bf 2},{\bf 1})$ of
$Spin(4)\times SU(2)_R$.   These scalars parameterize the ``Coulomb
branch.''   It is characterized
\ref\coulmet{S.J. Gates, C.M. Hull, M. Ro\v cek, Nucl. Phys. {\bf
B248} (1984) 157; G.W. Gibbons, G. Papadopoulos, K.S. Stelle,
hep-th/9706207.} 
by a generalized K\"ahler potential
$K(\Lambda,\Phi)$ satisfying the Laplace equation
\eqn\Laplace{K_{\Lambda\bar\Lambda}+K_{\Phi\bar\Phi}=0.}
It determines the metric 
\eqn\metric{ds^2=K_{\Phi\bar\Phi}d\Phi d\bar\Phi -
K_{\Lambda\bar\Lambda} d\Lambda d\bar\Lambda}
and the torsion
\eqn\fftor{B={1 \over 4} (K_{\Phi\bar\Lambda}d\Phi\wedge
d\bar\Lambda+K_{\Lambda\bar\Phi}d\bar\Phi\wedge d\Lambda).}
Note that the metric
is not hyper-K\"ahler.  The $\scrn=(4,4)$ supersymmetry is realized
in terms of three complex structures, which are covariantly constant
with respect 
to the generalized connection with torsion.  Unlike the Higgs branch,
here $SU(2)_R$ does not act on the space.  Instead, the $Spin(4)$
group acts on it and is spontaneously broken to a subgroup at a
generic point.

If the bosonic manifold defined by \metric\ admits a compact $U(1)$
isometry generated by the Killing vector
$i\left(\partial/\partial\Lambda-
\partial/\partial\bar\Lambda\right)$, we can dualize the twisted
superfield $\Lambda$ to an ordinary chiral superfield $\chi$.  This
breaks the $Spin(4)\times SU(2)_R$ invariance of the action to
$SU(2)_L\times SU(2)_R$ where $SU(2)_L \subset Spin(4)$ and there is
an extra $U(1)$ symmetry corresponding to the isometry. The
$\scrn=(4,4)$ sigma model with torsion is dual to a hyper-K\"ahler
sigma model with isometry group $SU(2)_L \times U(1)$ identical to the
one spanned by the hypermultiplets.  Note that manifest $\scrn=(4,4)$
invariance is preserved by this transformation, if and only if the
compact $U(1)$ isometry is {\it translational}, i.e.\ it preserves the
three complex structures.

The action may also contain bare parameters: mass terms for the matter
hypermultiplets transforming as $({\bf 2},{\bf 2},{\bf 1})$ of
$Spin(4)\times SU(2)_R$ and Fayet-Iliopoulos terms in the $({\bf
1},{\bf 1},{\bf 3})$ representation. Taking into account the 
$\scrn=(4,4)$ supersymmetry constraints and the transformation
properties under global symmetries we can prove certain
non-renormalization theorems analogous to those in \aps:
\item{1.} The gauge coupling constant can be promoted to a background
vector superfield, thus it can affect only the metric on the Coulomb
branch and not the Higgs branch.
\item{2.} The mass terms can also be regarded as scalar components of
vector superfields, thus they can affect only the Coulomb branch but
not the metric on the Higgs branch. 
\item{3.} The FI parameters can be promoted to hypermultiplets,
thus they can only appear in the metric on the Higgs branch.

At long distance the theory must be a superconformal field
theory\foot{Since the target space of the low energy effective theory
on the moduli space is noncompact, it is possible that the theory is
scale invariant but not conformally invariant.}.  The standard
$\scrn=4$ superconformal algebra includes an $SU(2)$ factor.
Therefore, we expect at long distance an $SU(2)\times SU(2)$ symmetry
under which the left moving supercharges are in $({\bf 2},{\bf 1})$
and the right moving supercharges in $({\bf 1},{\bf 2})$.  As the
theory on the Higgs branch is a sigma model on a hyper-K\"ahler
manifold, the $SU(2)\times SU(2)$ symmetry in the superconformal
algebra should not act on the bosons.  Therefore, we immediately
identify this $SU(2)\times SU(2)$ symmetry as the $Spin(4)$.  The
situation on the Coulomb branch is more complicated for two related
reasons.  First, the $Spin(4)$ symmetry acts on the bosons, and second
it is not a hyper-K\"ahler manifold.  In section 6 we will discuss the
superconformal algebra on the Coulomb branch and show that it includes
the $SU(2)_R$ symmetry.  Considerations similar to these have led
Witten
\refs{\separate,\wittenhiggs} 
to suggest that the Coulomb branch is decoupled from the Higgs
branch. 

\newsec{$A_{N_f-1}$ Theories -- $U(1)$ with $N_f$ electrons}

In this section we consider $U(1)$ gauge theories with $N_f$
electrons.  The global symmetry of these theories is $SU(N_f) \times
Spin(4) \times SU(2)_R$.  

For $N_f=0$ the theory is free.  For $N_f\geq 1$ the gauge
coupling, is a relevant operator.  The theory has a Coulomb branch
parameterized by the expectation values $\vec r\in R^4$ of the scalars
in the twisted multiplet.  The tree level metric is flat and the
torsion vanishes.  Equivalently,
\eqn\kclau{K_{cl}\left(\Phi,\bar\Phi,\Lambda,\bar\Lambda\right)=
{1\over g_{2}^2}(\Phi\bar\Phi-\Lambda\bar\Lambda).}

Quantum mechanically $K$ can be corrected.  In general, the geometry
of the target manifold is entirely determined by the function
\eqn\func{f\left(\Phi,\bar\Phi,\Lambda,\bar\Lambda\right)
=K_{\Phi\bar\Phi} =-K_{\Lambda\bar\Lambda}, }
which clearly satisfies the Laplace equation:
\eqn\lapeq{f_{\Phi\bar\Phi}+f_{\Lambda\bar\Lambda}=0.}
This equation has a unique $Spin(4)$ invariant solution given
by
\eqn\spinfso{f=A+{k \over \Phi\bar\Phi+\Lambda\bar\Lambda}}
for some constants $A$ and $k$.  (This fact is a special case of a
more general result in $\scrn= (0,4)$ supersymmetric theories
\ref\dps{M.R. Douglas, J. Polchinski and A. Strominger,
hep-th/9703031.}.)
We find it convenient to assign
dimension one to all the scalars (as they are components of gauge
fields in six dimensions).  Then the gauge coupling $g_2$ has
dimension one.  Therefore, $A$ in \spinfso\ can only appear at tree
level and $A= {1 \over g_2^2}$.  Similarly, $k$ can only
appear at one loop.  No higher order perturbative or non-perturbative
corrections of $K$ are possible.

An explicit one loop computation determines the value of $k=N_f$.
Therefore, the exact generalized K\"ahler potential is
\ref\exactK{M. Ro\v cek, K. Schoutens, A.Sevrin, Phys. Lett. {\bf
B265} (1991) 303; M. Ro\v cek, C. Ahn, K. Schoutens, A. Sevrin, 
hep-th/9110035, Contributed to Workshop on Superstrings and Related
Topics, Trieste, Italy, Aug 8-9, 1991,
Published in Trieste HEP Cosmol. (1991) 995.} 
\eqn\Ka{K\left(\Phi,\bar\Phi,\Lambda,\bar\Lambda\right)=
{1\over g_{2}^2}(\Phi\bar\Phi-\Lambda\bar\Lambda)-
N_f\int^{{\Lambda\bar\Lambda}\over{\Phi\bar\Phi}}
{d\xi\over \xi}\ln (\xi+1) +\ln\Phi\, \ln\bar\Phi.}
This leads to the metric
\eqn\meta{ds^2 = \left({1\over g_{2}^2}
+{N_f\over r^2}\right){d\vec r}^2}
and torsion
\eqn\Tora{B=-{1 \over 4} N_f \sin^2{\theta\over 2}d\phi\wedge d\chi}
where $0\leq\theta <\pi,\,0\leq\phi <2\pi,\,0\leq\chi <4\pi$
are angular coordinates on the unit three
sphere $S^3\subset R^4$
\eqn\angcoord{
\Phi= e^{i(\chi-\phi)/2}\cos{\theta\over 2},\qquad
\Lambda= e^{i(\chi+\phi)/2}\sin{\theta\over 2}.\qquad}
Strictly speaking, the two-form is not
globally defined on $S^3$, similarly to the Dirac monopole vector
potential, but it yields a well defined three-form field strength
proportional to the volume form of the sphere:
\eqn\threefo{H=-N_f d\Omega^3.}

It is possible to turn on bare masses $\vec m_i$, $i=1,\ldots,N_f$ for
the hypermultiplets (one of the masses can be set to zero by a choice
of the origin on the Coulomb branch).  The global $Spin(4)$ symmetry
is explicitly broken and the one-loop corrections on the Coulomb
branch are
\eqn\metb{\eqalign{
ds^2 &= \left({1\over g_{2}^2}+{1\over |\vec r-\vec m_1|^2}+ {1\over
|\vec r-\vec m_2|^2}+\ldots +{1\over |\vec r-\vec
m_{N_f}|^2}\right){d\vec r}^2\cr H &=
-d\Omega^{3}_1-\ldots-d\Omega^{3}_{N_f}}} where $\vec r = (\Phi,
\Lambda)$ and $d\Omega^{3}_i$ is the volume form on the unit sphere
centered around the point of coordinates $\vec m_i$.  The behavior of
$f$ near a particular point $\vec m_i$ is determined to be of the form
\spinfso\ by decoupling the remaining hypermultiplets ($\vec
m_j\rightarrow\infty,\ j\not =i$). This result is exact, since the
function $f\left(\Phi,\bar\Phi,\Lambda,\bar\Lambda\right)$ satisfies
the Laplace equation in $R^4\backslash\{\vec m_1,\ldots,\vec
m_{N_f}\}$ with fixed boundary conditions at infinity and around the
singular points $\vec m_i$.

In section 6 we will analyze the singularities in the metric in more
detail.  Here we simply note that the metric near the singularities
has an infinitely long tube. 

The Higgs branch of these models depends on the number of massless
flavors.  As explained in section 2, the metric on this space is
completely fixed at the classical level by a hyper-K\"ahler quotient
construction and is not corrected quantum mechanically.  For $N_f=1$
there is no Higgs branch.  For higher values of $N_f$ the Higgs branch
is isomorphic to the moduli space of a single $SU(N_f)$ instanton in
$R^4$.  The singularity at the origin corresponds to a point-like
instanton.

We now consider the corresponding $\scrn=4$ theories on $R^2 \times
S^1_R,$ where the circle is of radius $R$.  This should interpolate
between the three dimensional results of \refs{\IR,\swthree} and the
two dimensional theory.  In three dimensions electric-magnetic duality
transforms the vector multiplet into a hypermultiplet. The metric is
then constrained by symmetries and asymptotic one loop corrections to
be the Taub-NUT metric. Consider first the case where the radius $R$
is much larger than the distance scale set by the three dimensional
gauge coupling $g_3$.  Then, at energies above $1/R$ the theory is
three dimensional and the renormalization group evolution leads to the
Taub-NUT metric.  At energies below $1/R$ the renormalization group
evolution is that of the two dimensional sigma model based on this
space.  Since this sigma model is conformally invariant, it does not
evolve further and the Coulomb branch ends up being the Taub-NUT space
with an $A_{N_f-1}$ singularity (we expect the $\theta$ angle at the
singularity to vanish such that the 1+1 dimensional theory is not an
orbifold theory \aspinwall).  Note that one of the degrees of freedom
on this space is dual to the natural variables in the two dimensional
theory -- the space is parametrized by a hypermultiplet rather than by
a vector multiplet.  This is an explicit realization of the
speculation in the introduction.

We now compute the exact one loop corrected metric on $R^2 \times
S^1_R$ for any $R$.  The classical moduli space is $R^3 \times S^1$
where the $S^1$ factor is the Wilson line around the compact
dimension.  It is parameterized by the component of the three
dimensional gauge field $\sigma \sim \sigma + {1\over R}$.  The
three scalars in $R^3$ are denoted by $\vec \phi$.  Standard
dimensional reduction relates the three and the two dimensional
coupling constants
\eqn\coupling{{1\over g_{2}^2}={2\pi R\over g_{3}^2}.}
The relevant diagram is the vacuum polarization diagram with
hypermultiplets running in the loop
\eqn\sumonn{\sum_{n=-\infty}^{+\infty}\int {d^2p\over
{(2\pi)}^2 R}{1\over{\left(p^2+\vec \phi^2+
\left({n\over R}+{\sigma}\right)^2\right)^2}}.}
Using standard Poisson resummation techniques we find the one
loop corrected gauge coupling\foot{In the original version of this
paper the following equation had an error.  We thank J. Harvey and
G. Moore for pointing this out to us.}
\eqn\comp{{1\over g_{2}^2}+{N_f}{2\pi R\over |\vec \phi|}
\left\{{1\over 2}+\sum_{n\geq 1} e^{-2\pi R n|\vec\phi|}
\cos(2\pi Rn\sigma)\right\}.}
 
This result interpolates smoothly between the three dimensional and
the two dimensional regime. When the compactification radius is large
$R\gg{1\over |\vec \phi|}$ the exponential corrections are 
suppressed and the result reduces to the three dimensional one
\eqn\three{{1\over g_{2}^2}+{\pi R}{N_f\over |\vec\phi|}.}
For small radius $R\ll {1\over |\vec \phi|}$ the sum can be
approximated by an integral leading to the two dimensional result
\meta 
\eqn\two{{1\over g_{2}^2}+{N_f\over{{|\vec\phi|}^2+
{\sigma}^2}}.}

The answer \comp\ demonstrates the difficulty in proving the
speculation in the introduction.  In the three dimensional theory
$\sigma$ is a component of a vector field.  The Taub-NUT metric is
obtained after it is dualized to a scalar $\tilde \sigma$.  The three
dimensional theory has a global (magnetic) $U(1)$ symmetry under which
$\tilde \sigma$ is shifted by a constant.  If we dimensionally reduce
the tree level theory, then $\sigma$ and $\tilde \sigma$ are related
by ordinary T-duality.  However, the one loop corrected metric \comp\
shows that the compactified theory no longer has a continuous symmetry
shifting $\sigma$ by a constant.  Therefore, it is not clear how to
dualize it.  Similar problems of T-duality with respect to
transformations, which are not symmetries, are familiar in two
dimensions
\ref\morple{D. Morrison and R. Plesser, hep-th/9508107, Contributed to
STRINGS 95: Future Perspectives in String Theory, Los Angeles, CA,
13-18 Mar 1995.}.
 
In the three dimensional theory every mass term has three real
components in $({\bf 3},{\bf 1})^0$ of the global $SU(2)_L\times
SU(2)_R\times U(1)$.  After the compactification to two dimensions
there are four real parameters.  The fourth one arises as a Wilson
line of the corresponding global symmetry around the circle.  More
generally, starting in six dimensions and compactifying on a four
torus, all four mass parameters arise as Wilson lines.  With non-zero
masses in three dimensions and Wilson lines the one loop result
becomes
\eqn\masses{{1\over g_{2}^2}+\sum_{i=1}^{N_f}{2\pi R\over
|\vec\phi-\vec m_i|}\left\{{1\over 2}+\sum_{n\geq 1}
e^{-2\pi Rn|\vec\phi-\vec m_i|}
\cos[2\pi Rn(\sigma-\sigma_i)]\right\}.}

This two dimensional field theory appears as the effective theory on a
D1-brane near $N_f$ D5-branes.  The $U(1)$ vector multiplet arises
{}from strings, whose two ends touch the D1-brane, and the $N_f$
electrons arise from strings, which connect the D1-brane to the
D5-branes.  The Coulomb branch corresponds to the situation, where the
D1-brane is separated from the D5-branes but is parallel to them.  The
metric \meta\ is the standard metric near a D5-brane obtained by
solving the classical supergravity equations \chs.  Similarly, the
torsion \Tora\ represents the $B$ field generated by the charge of the
D5-branes.  

We can also study a single D2-brane and $N_f$ D6-branes wrapping a 
circle of radius $R$.  The effective theory corresponding to this
system is the three dimensional gauge theory compactified on a circle
which we discussed above.

\newsec{$D_{N_f}$ Theories --$SU(2)$ with $N_f$ flavors}

Here we study $SU(2)$ gauge theories with $N_f$ quark flavors in the
fundamental representation.  The global symmetry of these theories is
$SO(2N_f) \times Spin(4) \times SU(2)_R$.  The analysis of these
theories is similar to the discussion in the previous section with
some minor differences.
 
Unlike the $U(1)$ theories the Coulomb branch is not $R^4$ but
$R^4/Z_2$ where the $Z_2$ originates from the Weyl group of $SU(2)$.
The classical metric on this space is flat.
As in the $U(1)$ theories, the global $Spin(4)$ symmetry completely
determines the quantum metric up to one coefficient which is easily
computed at one loop 
\eqn\metc{ds^2=\left({1\over g_{2}^2}+{2N_f-2\over
r^2}\right){d\vec r}^2}
(the $-2$ represents the contribution of the non-Abelian vector
multiplet).  The corresponding torsion is 
\eqn\torb{H=-2(N_f-1)d\Omega^{3}.}
 
The cases $N_f=0,\,1$ are particularly interesting.  For $N_f=0$ the
metric has a singularity at finite distance in the moduli space
$r_0=g_{2}\sqrt{2}$ and becomes negative definite past this value of
$r$. This signals the fact that the present variables are not adequate
to describe the physics in that region.  For $N_f=1$ the effects of
the charged hypermultiplet cancel precisely those of the non-Abelian
vector multiplet and the metric remains flat and there is no tube in
the metric. Similar phenomena occur in $SU(2)$ gauge theory with
$N_f=2^{d-2}$ in $d=3,4,5,6$ dimensions.

Adding masses for hypermultiplets explicitly breaks the $Spin(4)$
symmetry and results in quantum corrections of the form
\eqn\metd{\eqalign{
ds^2 &=\left({1\over g_{2}^2}+{1\over |\vec r-\vec m_1|^2}
+{1\over |\vec r+\vec m_1|^2}+\ldots+{1\over |\vec r-\vec m_{N_f}|^2}
+{1\over |\vec r+\vec m_{N_f}|^2}-{2\over r^2}\right)d\vec r^2\cr
H &= -d\Omega^3_{1-}-d\Omega^3_{1+}-\ldots 
     -d\Omega^3_{N_f-}-d\Omega^3_{N_f+}+2d\Omega^3\cr}} 
where $d\Omega^{3}_{i\pm}$ is the volume form on the unit sphere
centered around the point of coordinates $\pm \vec m_i$.  As before,
this metric is exact.

The Higgs branch is again determined at the classical level by
hyper-K\"ahler quotient constructions and is not corrected quantum
mechanically.  It is isomorphic to the moduli space of a single
$SO(2N_f)$ instanton in $R^4$ (there is no Higgs branch for
$N_f=0,1$).

In three dimensions the Coulomb branch of these theories can
be described in hypermultiplet variables.  For $N_f\geq 2$ the metric
is determined at one loop to be Taub-NUT \swthree. For $N_f=0,1$ there
are instanton corrections resulting in the Atiyah-Hitchin metric
\swthree.  Reduction to two dimensions on a large circle yields an
$\scrn= (4,4)$ sigma model with the corresponding hyper-K\"ahler
metric.  As in the $U(1)$ case, this metric leads to a scale invariant
two dimensional field theory with $D_{N_f}$ singularity (we expect the
$\theta$ angle at the singularity to vanish such that the 1+1
dimensional theory is not an orbifold theory \aspinwall).  Again, we
see an explicit realization of the speculation in the introduction.
For the special cases $N_f=0,1$ the metric is not singular and the
problem of the sign of the metric for $N_f=0$ is absent.
Unfortunately, we do not know how to extend this discussion to smaller
$R$ essentially for the reasons explained in the previous section (the
compactified theory does not have the global symmetry needed for the
T-duality).

The brane realization of these $SU(2)$ theories is achieved on a
D1-brane near $N_f$ D5-branes and an orientifold.  The asymptotic
transverse space is $R^4/Z_2$ and the one loop corrected metric \metc\
is the spacetime metric obtained by solving the classical supergravity
equations.  Similarly, the torsion \torb\ represents the $B$ field
generated by the charge of the orientifold and the $N_f$ D5-branes --
charge $-2$ from the orientifold and $+2N_f$ from the D5-branes.
Again, the Coulomb branch corresponds to the D1-brane away from the
D5-branes and the orientifold and the Higgs branch represents the
absorption of the D1-brane in them as an instanton.

\newsec{$E_{6,7,8}$}

In higher dimensions there exist other theories with a one dimensional
Coulomb branch.  Their global symmetries are $E_{6,7,8}$.  They do not
have $E_{6,7,8}$ invariant 
Lagrangians, which flow to them\foot{In three dimensions they can be
found in the long distance limit of local field theories based on 
Lagrangians, which are not $E_{6,7,8}$ invariant
\ref\intse{K. Intriligator and N. Seiberg, Phys.Lett. {\bf B387}
(1996) 513, hep-th/9607207.}.}.  
When we compactify them on tori to two dimensions we should
find two dimensional theories with these global symmetries.  We take
such a compactification as the definition of the two dimensional
theories. 

The Higgs branches of these theories in higher dimensions are
isomorphic to the moduli spaces of a single $E_{6,7,8}$ instanton in
$R^4$.  Since the parameters of the compactification are in vector
multiplets, they cannot affect the Higgs branches.  Therefore, the
Higgs branches of these theories in two dimensions are also isomorphic
to the moduli spaces of an $E_{6,7,8}$ instanton in $R^4$.

Since we do not have a Lagrangian description of these theories, we
cannot analyze the region far out along the Coulomb branch.  In the
next section we will describe these theories near the origin.

In string theory these theories occur on a D1-brane probing a
background of 5-branes at strong coupling.

\newsec{The Theory in The Tube}

In this section we consider the region in the Coulomb branch around
the origin.  If we consider a fixed region around a point $\vec r$
in the moduli space and take the strong coupling limit $g_2
\rightarrow \infty$ (recall that we define the fields such that there
is a factor of $1 \over g_2^2$ in front of the action), the metric and
torsion become
\eqn\metE{\eqalign {& ds^2=k{d\vec r^2\over r^2}\cr
                    & H=-kd\Omega^{3}\cr}}
where $k$ denotes the total $B$-field charge 
\eqn\charge{k=\left\{\matrix{ 
N_f &\qquad\hbox{for \ }\ {U(1)}\hfill\cr
2(N_f-1) &\qquad\hbox{for\ }\ SU(2).\cr}\right.} 

For $SU(2)$ with $N_f=1$, $k=0$, and therefore the metric on the
Coulomb branch is flat and proportional to $1 \over g_2^2$.  For
$SU(2)$ with $N_f=0$ the metric we find is not Riemannian, and
therefore there must be a new description of the theory.

For $k\geq 1$, a change of variables 
$\phi=\sqrt{k}ln\sqrt{r^2/k}$ yields the familiar tube \chs 
\eqn\throat{\eqalign
{& ds^2=d\phi^2+kds_3^2,\cr
& H=-kd\Omega^{3}}}
where $ds_3^2$, $d\Omega^{3}$ are the metric and the volume form on
the unit three sphere $S^3\subset R^4$.  The metric and anti-symmetric
tensor field on $S^3$ define a level $k$ supersymmetric $SU(2)$ WZW
model and the non-trivial dilaton defines a supersymmetrized
Liouville field with background charge
\eqn\bkcharge{Q=\sqrt{2\over k}.}
For the $D_{N_f}$ theories the WZW model is defined on the group
manifold $SO(3) \sim S^3/Z_2$ rather than on its double cover,
reflecting the $Z_2$ action on the Coulomb branch. This is consistent
with the level of the corresponding affine algebra being even.

The supersymmetric WZW theories are analyzed by first performing a
chiral rotation making the fermions free.  This has the effect of
shifting the coefficient of the WZ term for the bosons $k\rightarrow
\tilde k = k-2$.  One way to understand this is to note that the three
free fermions lead to an $\widehat{ SU(2)}_2$ algebra and the bosons
to $\widehat{ SU(2)}_{\tilde k=k-2}$, which is generated by $J_i$.
The diagonal affine algebra has level $\tilde k +2 = k$.
Therefore, for $k>2$ we are left with a WZW theory with level $\tilde
k >0$.  For the two cases with $k=2$ (the $A_1$ and the $D_2$
theories) there is no bosonic WZW theory -- the resulting theory has
only the boson $\phi$ in \throat\ and four free fermions.  For $k=1$
(the $A_0$ theory) $\tilde k <0$.  Therefore, the supersymmetric WZW
theory is not unitary and a new description of the theory should be
found.

The four free fermions $\psi^a$ ($a=0$ for the superpartner of
$\phi$ and $a=1,2,3$ for the free fermions obtained by the chiral
rotation ) lead to an $\widehat{ SU(2)}_1 \times \widehat{ SU(2)}_1$
algebra generated by 
\eqn\suttim{j_i^\pm=\epsilon_{ijk}\psi^j \psi^k  \mp \psi^0 \psi^i}
(the previously mentioned $\widehat{ SU(2)}_2$ is the diagonal
subalgebra).  At this stage we can compare the various symmetries with
the classical $Spin(4) \times SU(2)_R$ symmetry of our theory.  From
the action of the $Spin(4) \cong SU(2) \times SU(2)$ on the
supercharges it is clear that one $SU(2)$ leads to a left moving
symmetry and the other $SU(2)$ to a right moving 
symmetry.  Since it acts on the bosons, the corresponding current must
include $J_i$.  The anomaly in this current can be computed
semiclassically \wittenhiggs.  The fermions in the hypermultiplets
contribute $1$ and the fermions in the vector multiplets contribute
$-1$.  Therefore, the anomaly is $N_f-1=k-1$ in the $A_{N_f-1}$
theories and $2N_f-3=k-1$ in the $D_{N_f}$ theories.  The $SU(2)_R$
symmetry does not act on the bosons on the Coulomb branch.  Therefore,
it must be extended in the long distance theory to two $SU(2)$
symmetries -- one for the left movers and one for the right movers.
The four right moving supercharges should therefore be in $({\bf
2,2})$ of an $SU(2)$ from the $Spin(4)$ and a new $SU(2)$ which does
not act on the bosons.  This together with the anomalies leads to the
identification of these two $\widehat{ SU(2)}$ as being generated by
\eqn\KMgen{\eqalign{
& A_i^{+}=j_i^++J_i \cr
& A_i^{-}=j_i^- . \cr}}

Using these currents and the energy momentum and supercurrents
\eqn\susycharges{\eqalign{
&T=-J^0 J^0-{1\over k}J^i J^i-\partial\psi^a\psi^a \cr
& G^0=2\left\{J^0\psi^0+{1\over\sqrt{k}}J^i\psi^i+
{2\over \sqrt{k}}\psi^1\psi^2\psi^3\right\}\cr
& G^1=2\left\{J^0\psi^1+{1\over\sqrt{k}}\left(
-J^1\psi^0+J^2\psi^3-J^3\psi^2\right)-
{2\over\sqrt{k}}\psi^0\psi^2\psi^3\right\}\cr}}
($J^0$ is the current of the Liouville field $\phi$) and cyclic
permutations for $G^2,\ G^3$ we can find \chs\ the extended $\scrn=4$
superconformal algebra with $SU(2)\times SU(2)\times U(1)$ symmetry
\ref\large{A. Sevrin, W. Troost, A. van Proeyen, Phys. Lett. {\bf
B208} (1988) 601.}.
For $\tilde k=0$ ($k=2$) the bosonic WZW theory is trivial
and we should set $J^{1,2,3}=0$ in \susycharges.  The effect of the
background charge \bkcharge\ can be taken into account by redefining
the energy-momentum tensor and supersymmetry charges
\eqn\improved{
\tilde T = T-{1\over \sqrt{k}}\partial J^0,\qquad 
\tilde G^a=G^a-{1\over\sqrt{k}}\partial\psi^a.}
Now $\tilde T$, $\tilde G^a$ and the level one  
currents $A_i^{-}$ satisfy the standard $\scrn=4$ algebra with $c=6$.

Conformal field theories are characterized not only by their chiral
algebra but also by forming a modular invariant using the
representations of the chiral algebra.  In our case we should find the
modular invariants of $\widehat{ SU(2)}_{\tilde k}$.  These are given
by an $A-D-E$ classification \ciz
\eqn\modinv{
\matrix{
A_{\tilde k+ 1}\qquad & \tilde k\geq 0 \hfill \cr
D_{{\tilde k\over 2}+2}\qquad & \tilde k\geq 4,\ \tilde k\ \hbox{even}
\cr 
E_6\hfill\qquad  &\tilde k=10 \hfill \cr
E_7\hfill\qquad &\tilde k=16 \hfill \cr
E_8\hfill\qquad &\tilde k=28 .\hfill  \cr}}
In all cases $\tilde k = k-2=h(G)-2$, where $h(G)$ is the dual
Coxeter number of the group $G=A_{\tilde k +1},\ D_{{\tilde k \over 2}
+2}, \ E_{6,7,8}$.  The $A_{\tilde k+ 1}$ theories are obtained in WZW
models on the $SU(2)$ group manifold with level $\tilde k$ and the
$D_{{\tilde k\over 2}+2}$ theories are obtained in WZW theory on the
$SO(3)=SU(2)/Z_2$ group manifold with level $\tilde k$ 
\ref\gepwit{D. Gepner and E. Witten, Nucl. Phys. {\bf B278} (1986)
493.}.

In our case the $U(1)$ gauge theories with $N_f$ electrons
($A_{N_f-1}$) lead to the standard WZW theories on the $SU(2)$ group
manifold with level $\tilde k = N_f-2$ with its diagonal modular
invariant ($A_{\tilde k+1}$ in \modinv).  Our $SU(2)$ gauge theories
with $N_f$ quarks ($D_{N_f}$) lead to a WZW theory on the
$SO(3)=SU(2)/Z_2$ group manifold with level $\tilde k = 2N_f-4$ and
hence to the $D_{{\tilde k\over 2}+2}$ modular invariant in \modinv.
Finally, in section 5 we mentioned three more theories $E_{6,7,8}$.
They are naturally identified with the $E_{6,7,8}$ modular invariants
in \modinv.

This provides an explanation of the well known but somewhat mysterious
relation between $A-D-E$ groups and modular invariants of $\widehat
{SU(2)} $.  Our 
theories with $A-D-E$ global symmetries lead to an $\widehat
{SU(2)}_{\tilde k}$ algebra with $A-D-E$ modular invariants.  This is
similar to the result of \refs{\IR,\swthree}, where the same theories
in three dimensions were studied.  There a relation between the
$A-D-E$ global symmetries and the singularities in the Coulomb
branches was found, thus explaining the relation between the $A-D-E$
groups and $A-D-E$ singularities and discrete subgroups of $SO(3)$ (a
closely related relation in the context of string theory had been
found in
\ref\witdyn{E. Witten, Nucl.Phys. {\bf B443} (1995) 85,
hep-th/9503124.}).
Finally, if the speculation in the introduction is correct, there can
be a direct relation between all these three places where an $A-D-E$
classification exists.  The theories with $A-D-E$ global symmetry in
two dimensions have a Coulomb branch with an $A-D-E$ modular invariant
and a dual description similar to the one in three dimensions with an
$A-D-E$ singularity.

We should restate that for $SU(2)$ with $N_f=0,1$ ($D_{0,1}$) and for
$U(1)$ with $N_f=1$ ($A_0$) the description in terms of the tube
breaks down.  Precisely in these cases there is no Higgs branch, and
correspondingly in the string applications there is no enhanced
symmetry.  It seems that the Coulomb branch for $D_1$ is simply
$R^4/Z_2$.  It will be interesting to understand the other cases.

\centerline{\bf Acknowledgments}\nobreak

We would like to thank P. Aspinwall, T. Banks, M. Douglas, J. Harvey,
K.  Intriligator, G. Moore, S. Shenker and E. Witten for useful
discussions.  This work was supported in part by DOE grants
\#DE-FG02-90ER40542 and \#DE-FG02-96ER40559. 

\listrefs
\end